\documentstyle[prd,aps]{revtex}

\input epsf
\epsfverbosetrue

\def\Journal#1#2#3#4{{#1} {\bf #2} (#4) #3}

\def\SCI{Science}
\def\APJ{Astrophysics J.}
\def\MPL{Mod. Phys. Lett. A}

\def\NPB{Nucl. Phys. B}
\def\NPBOLD{Nucl. Phys.}

\def\PLB{{Phys. Lett.} B}
\def\PREPO{Phys. Rep.}
\def\PLBOLD{Phys. Lett.}
\def\PRL{Phys. Rev. Lett.}
\def\RMP{Rev. Mod. Phys.}
\def\PRD{Phys. Rev. D}

\def\PTP{Prog. Theor. Phys.}
\def\JHEP{JHEP}

\def\JETPUSSR{JETP (USSR)}
\def\ZETP{Zh. Eksp. Teor. Piz.}
\def\TNYAS{Trans. New York Acad. Sci.}
\def\IJMP{Int. J. Mod. Phys. A}

\def\mapgeq{\mathbin{\lower.3ex\hbox{$\buildrel>\over{\smash{\scriptstyle\sim}\vphantom{_x}}$}}}
\def\mapleq{\mathbin{\lower.3ex\hbox{$\buildrel<\over{\smash{\scriptstyle\sim}\vphantom{_x}}$}}}
\def\mapgeqeq{\mathbi{\lower.3ex\hbox{$\buildrel>\over{\smash{\scriptstyle\approx}\vphantom{_2}}$}}}
\def\mapleqeq{\mathbin{\lower.3ex\hbox{$\buildrel<\over{\smash{\scriptstyle\approx}\vphantom{_2}}$}}}

 \mathchardef\#="0023
 \mathchardef\$="0024
 \mathchardef\%="0025
 \mathchardef\ddash="705C
 
 \mathchardef\lwavy="336E
 \mathchardef\rwavy="336F
 \mathchardef\biglwavy="331A
 \mathchardef\bigrwavy="331B
 \mathchardef\bigglwavy="3328
 \mathchardef\biggrwavy="3329
 \mathchardef\littlesum="0350

\tighten
\draft
\begin{document} 
\bibliographystyle{prsty}

\title{
The Interplay between Neutrinos and Charged Leptons\\
in the Minimal $SU(3)_L\times U(1)_N$ Gauge Model
}

\author{
Teruyuki Kitabayashi$^a$
\footnote{E-mail:teruyuki@post.kek.jp}
and Masaki Yasu${\grave {\rm e}}^b$
\footnote{E-mail:yasue@keyaki.cc.u-tokai.ac.jp}
}

\address{\vspace{5mm}$^a$
{\sl Accelerator Engineering Center} \\
{\sl Mitsubishi Electric System \& Service Engineering Co.Ltd.} \\
{\sl 2-8-8 Umezono, Tsukuba, Ibaraki 305-0045, Japan}
}
\address{\vspace{2mm}$^b$
{\sl Department of Natural Science\\School of Marine
Science and Technology, Tokai University}\\
{\sl 3-20-1 Orido, Shimizu, Shizuoka 424-8610, Japan\\and\\}
{\sl Department of Physics, Tokai University} \\
{\sl 1117 KitaKaname, Hiratsuka, Kanagawa 259-1292, Japan}}
\date{TOKAI-HEP/TH-0103, March, 2001}
\maketitle

\begin{abstract}
In the minimal $SU(3)_L\times U(1)_N$ gauge model with a global $L_e-L_\mu-L_\tau$ ($=L^\prime$) symmetry and a discrete $Z_4$ symmetry, it is found that the interplay between neutrinos and charged leptons contained in triplets of $\psi^i$=($\nu^i_L$, $\ell^i_L$, $\ell^{ci}_L$) ($i$=1,2,3) naturally leads to the large mixing angle (LMA) MSW solution.  The model includes two (anti)sextet Higgs scalars, $S^{(0)}$ with $L^\prime$=0 and $S^{(+)}$ with $L^\prime$=2, which, respectively, couple to $\psi^1\psi^{2,3}$ for the electron mass and masses of atmospheric neutrinos and to $\psi^{2,3}\psi^{2,3}$ for the $\mu$- and $\tau$-masses and one-loop radiative neutrino masses relevant to solar neutrinos.  This mechanism is realized by utilizing an additional residual discrete symmetry supplied by explicitly broken $L^\prime$, which guarantees the absence of tree-level neutrino mass terms of the  $\psi^{2,3}\psi^{2,3}$-type.  Pure rotation effects due to the diagonalization of neutrino and charged-lepton mass matrices are estimated to yield $\Delta m^2_\odot/\Delta m^2_{atm}\mapleq (m_e/m_\mu)^{3/2}={\mathcal{O}}(10^{-4})$ but the radiative effects supersede the rotation effects to yield $\Delta m^2_\odot/\Delta m^2_{atm}={\mathcal{O}}(10^{-2})$ as the LMA solution.

\end{abstract}
\vspace{2mm}
\section{Introduction}
In the recent report on the detailed analysis of solar neutrino oscillations \cite{OldSolarTh,OldSolar,Solar} done by the Super-Kamiokande collaboration \cite{RecentSK}, it has been suggested that solutions with large mixing angles are favored while solutions with small mixing angles are disfavored at the 95$\%$ confidence level.  Neutrino oscillations controlled by such large mixing angle have also been observed and confirmed for atmospheric neutrinos \cite{Kamiokande}.  Both observed oscillations are characterized by the same property of neutrino mixings, namely, large neutrino mixings, which are theoretically consistent with bimaximal mixing scheme \cite{Mixing,NearlyBiMaximal}.  The difference arises in their oscillation scales denoted by $\Delta m^2_{atm}$ for atmospheric neutrinos and $\Delta m^2_\odot$ for solar neutrinos, which are specified by $\Delta m^2_{atm}\sim 3\times 10^{-3}$ eV$^2$ and $\Delta m^2_\odot\mapleq 10^{-4}$ eV$^2$, thereby, indicating a hierarchy of $\Delta m^2_{atm}\gg\Delta m^2_\odot$.

Neutrino oscillations arise for massive neutrinos \cite{EarlyMassive}.  Since neutrino masses of $\sim 5.5\times 10^{-2}$ eV are implied by $\Delta m^2_{atm}$, models for neutrino oscillations must be equipped with mechanisms generating such tiny masses of neutrinos  \cite{SeeSaw,1-loop,2-loop}, which include breaking of the lepton number conservation for Majorana neutrino masses.  It has been long known \cite{SU3UNuMass} that such lepton number breaking interactions are inevitably contained in the minimal $SU(3)_L \times U(1)_N$ model for electroweak interactions (called the 331 model) \cite{SU3U1} because a charged lepton ($\ell$) and its antiparticle ($\ell^c$) together with a neutrino ($\nu$) are placed in a triplet of $SU(3)_L$ as $\psi$=($\nu_L$, $\ell_L$, $\ell^c_L$)$^T$.  As a result, the model provides a charged-lepton mass term included in $\psi\psi$, which simultaneously contains $\nu_L\nu_L$ as a neutrino mass term. Namely, the need for Majorana mass term is linked to the existence of the massive charged leptons. Therefore, the 331 model is suitable for incorporating Majorana neutrino masses \cite{SU3UNuMass,KitaYas,NuVEV,VeryLately}.

With this nice feature of the 331 model in mind, we introduce the bimaximal mixing scheme into the model to accommodate observed neutrino oscillations.  One of the possible scenario for bimaximal mixing scheme is to employ a new $U(1)_{L^\prime}$ symmetry based on $L_e-L_\mu-L_\tau$ ($\equiv$ $L^\prime$) \cite{EarlierLprime,Lprime}.  Neutrino mixings are predicted to be maximal for solar neutrinos with $\Delta m^2_\odot$=0 but arbitrary for atmospheric neutrinos with $\Delta m^2_{atm} \neq 0$.  The hierarchy of $\Delta m^2_{atm}\gg\Delta m^2_\odot$ is trivially satisfied since $\Delta m^2_\odot$=0.  One has to next explain how a nonvanishing $\Delta m^2_\odot$ is induced while preserving $\Delta m^2_{atm}\gg\Delta m^2_\odot$.  An elaborated mechanism is to generate $\Delta m^2_\odot$ as radiative effects \cite{tree-1loop-2loop}, whose generic smallness explains the relative smallness of $\Delta m^2_\odot$ over $\Delta m^2_{atm}$ \cite{1loop_and_or2loop,SUSY_tree-1loop}.

In this article, we construct our interactions that induce phenomenologically consistent neutrino and charged-lepton mass matrix, $M_\nu$ and $M_\ell$, with appropriate radiative effects for $\Delta m^2_\odot$.  Since the model describes both matrices in a unified form, masses for charged leptons and neutrinos are correlated to each other.  The mass term of $\psi\psi$ transforms as $3^\ast$ and $6$, which requires $3$ and/or $6^\ast$ as Higgs scalars.  The coupling of the triplet Higgs scalar to $\psi\psi$ is antisymmetrized with respect to the $SU(3)_L$-index. The spin and statistics, then, requires antisymmetrized flavor indices carried by $\psi$, in turn, generating antisymmetrized $M_\ell$ with eigenvalues corresponding to $m_\mu=m_\tau$, which contradicts with the observed hierarchy of $m_\mu \ll m_\tau$. The triplet Higgs scalar alone is not consistent with the observed lepton mass spectrum.  Therefore, the sextet scalar, which provides symmetrized mass terms, is the key ingredient of the 331 model \cite{Sextet}.
\footnote{
The use of an $SU(3)_L$-singlet vector-like heavy $\ddash$electron" without introducing a sextet scalar \cite{HeavyElectron} has been advocated to describe consistent mass spectra of neutrinos and charged leptons \cite{VeryLately}.}

Because the sextet Higgs scalar contains both an $SU(2)_L$-doublet scalar for charged leptons and an $SU(2)_L$-triplet scalar for neutrinos, we naturally expect $\ddash$degeneracy" in the structure of their mass matrices.  The charged leptons should contain a diagonal mass term such as the $\tau$-mass, which forces $M_\nu$ to have a diagonal term.  However, $M_\nu$ should not contain diagonal terms to be consistent with bimaximal mixing scheme based on $L^\prime$.  To avoid this $\ddash$degeneracy", we use the conservation of the $L^\prime$ quantum number supplemented by an appropriate $Z_4$ parity as well as a specific breaking pattern of $L^\prime$ yielding a residual discrete symmetry.  The collaboration of these symmetries allows neutrinos to develop no diagonal mass terms.  As a result, atmospheric neutrino oscillations are controlled by those tree-level off-diagonal masses while solar neutrino oscillations are to be induced by one-loop effects.
\footnote{
Similar discussions based on $L^\prime$ and a discrete symmetry have been done in Ref.\cite{TreeOneLoop}.
}

In the following section, we first introduce particles and their interactions and discuss how the required vacuum alignment is realized. Section III deals with the diagonalization of $M_\nu$, which reflects rotation effects due to the diagonalization of $M_\ell$.  The bimaximal condition on the neutrino masses determines a pattern of the matrix elements for the electron mass.  One-loop radiative mass of $\nu_e$ is also calculated.  The discussions in Sec.IV include to examine $\Delta m^2_\odot$ by utilizing the correlation between $M_\nu$ and $M_\ell$ and to confirm that the partial $\ddash$degeneracy" between $M_\nu$ and $M_\ell$ suppresses the rotation effects to yield the LMA solution based on the radiative effects.  The final section is devoted to summary and  discussions.

\section{Model}
The model contains neutrinos and charged leptons in triplets of $\psi^i_\alpha=(\nu^i_L, \ell^i_L, \ell^{ci}_L)^T$ ($i,\alpha$=1,2,3)
\footnote{
Throughout this paper, the roman letters of $i,j,\cdots$ and the Greek letters of $\alpha, \beta, \cdots$, respectively, stand for the three families and three $SU(3)_L$-indices.
}
 with $L^\prime=1$ for $\psi^1$ and $L^\prime=-1$ for $\psi^{2,3}$, where the superscript $c$ of $\ell^{ci}_{\alpha L}$ stands for the charge conjugation, and two (anti-)sextet Higgs scalars $S^{(0,+)\alpha \beta}$ with $L^\prime$=(0, 2) for ($S^{(0)}$, $S^{(+)}$).  Since $S^{(0)}$ can couple to $\psi^1\psi^i$ ($i$=2,3), $\nu_{eL} - \nu^i_L$ mixing terms are generated by $\langle 0 \vert S^{(0)11} \vert 0\rangle\neq 0$ and are to be responsible for the bimaximal structure.  On the other hand, $m_\tau$ (and also $m_\mu$) are generated by $\langle 0 \vert S^{(+)23} \vert 0\rangle\neq 0$ via its coupling to $\psi^i\psi^j$ ($i,j$=2,3).  Since $\psi^i\psi^j$ also supply couplings to $\nu_{\mu L}$ and $\nu_{\tau L}$, if $\langle 0 \vert S^{(+)11} \vert 0\rangle \neq 0$ is also induced, then it disturbs the nice feature of the bimaximal mixing scheme based on $L^\prime$.  Therefore, in order for the present approach to be acceptable, the simultaneous alignment of $\langle 0 \vert S^{(+)11} \vert 0\rangle=0$ and $\langle 0 \vert S^{(+)23} \vert 0\rangle\neq 0$ should be dynamically allowed.  It is demonstrated that this vacuum alignment of $S^{(+)11}$ and $S^{(+)23}$ is indeed possible to occur as a result of the emergence of a residual discrete symmetry specific to $U(1)_{L^\prime}$.

\vspace{-5mm}
\subsection{Particles}
The 331 model is specified by the $U(1)_N$-charge.  Let $N/2$ be the $U(1)_N$ quantum number, then the hypercharge, $Y$, is given by $Y=-\sqrt{3}\lambda^8+N$ and the electric charge $Q_{em}$ is given by $Q_{em}=(\lambda^3+Y)/2$, where $\lambda^a$ is the $SU(3)$ generator with Tr$(\lambda^a \lambda^b)=2\delta^{ab}$ $(a,b=1...8)$.  Let us first summarize the whole particle content in our model, where the quantum numbers are specified by ($SU(3)_L$, $U(1)_N$; $U(1)_{L^\prime}$):
\begin{equation}
-\psi^1=\left(\nu^1_L, \ell^1_L, \ell^{c1}_L\right)^T : \left( {\bf 3}, 0; 1 \right), \quad
\psi^{i=2,3}=\left(\nu^i_L, \ell^i_L, \ell^{ci}_L\right)^T  : \left( {\bf 3}, 0; -1 \right), 
\label{Eq:Leptons}
\end{equation}
for leptons,
\begin{eqnarray}
&
Q^{i=1,2}_L=\left(d^i_L, -u^i_L, J^i_L\right)^T : \left( {\bf 3}^\ast, -1/3; 0 \right),
\quad
Q^3_L=\left(u^3_L, d^3_L, J^3_L\right)^T : \left( {\bf 3}, 2/3; 0 \right), 
\nonumber \\
&u^{1,2,3}_R : \left( {\bf 1}, 2/3; 0\right),\quad 
d^{1,2,3}_R : \left( {\bf 1},-1/3; 0  \right),\quad 
J^{1,2}      _R : \left( {\bf 1}, -4/3; 0  \right),\quad 
J^3_R : \left( {\bf 1}, 5/3; 0  \right),
\label{Eq:Quarks}
\end{eqnarray}
for quarks
\begin{eqnarray}
& \eta=\left(\eta^0, \eta^-, \eta^+\right)^T : \left( {\bf 3}, 0; 0 \right), \quad
   \rho=\left(\rho^+, \rho^0, \rho^{++}\right)^T : \left( {\bf 3}, 1; 0 \right), \quad
   \chi=\left(\chi^-, \chi^{--}, \chi^0\right)^T : \left( {\bf 3}, -1; 0 \right),
\label{Eq:HiggsTriplet}
\end{eqnarray}
for triplet Higgs scalars, and
\begin{eqnarray}
&  S^{(0)}=
\left( {\begin{array}{*{20}c}
   {s^{(0)}_\nu  } & {s^ {(0)+}  } & {s^ {(0)-}  }  \\
   {s^{(0)+}  } & {s^{(0) +  + } } & {s^{(0)}_\ell  }\\
   {s^{(0)-}  } & {s^{(0)}_\ell  } & {s^{(0) -  - } }  \\
\end{array}} \right) : \left( {\bf 6^\ast}, 0; 0 \right), \quad
  S^{(+)}=
\left( {\begin{array}{*{20}c}
   {s^{(+)}_\nu  } & {s^ {(+)+}  } & {s^ {(+)-}  }  \\
   {s^{(+)+}  } & {s^{(+) +  + } } & {s^{(+)}_\ell  }\\
   {s^{(+)-}  } & {s^{(+)}_\ell  } & {s^{(+) -  - } }  \\
\end{array}} \right) : \left( {\bf 6^\ast}, 0; 2 \right), 
\label{Eq:HiggsSextet}
\end{eqnarray}
for (anti-)sextet Higgs scalars. These Higgs scalars have the following vacuum expectation values (VEV's):
\begin{eqnarray}
&\langle 0 \vert \eta\vert 0 \rangle =\left(v_\eta, 0, 0\right)^T, \quad 
\langle 0 \vert \rho\vert 0 \rangle  =\left(0, v_\rho, 0\right)^T, \quad 
\langle 0 \vert \chi\vert 0 \rangle =\left(0, 0, v_\chi\right)^T,
\label{Eq:VEV_Triplet} \\
&  \langle 0 \vert S^{(0)}\vert 0 \rangle =
\left( 
\begin{array}{*{20}c}
   v^{(0)}_\nu & 0 & 0 \\
   0 & 0 & v^{(0)}_\ell\\
   0 & v^{(0)}_\ell & 0 \\
\end{array}
\right),
\quad
  \langle 0 \vert S^{(+)}\vert 0 \rangle =
\left( 
\begin{array}{*{20}c}
   0 & 0 & 0 \\
   0 & 0 & v^{(+)}_\ell\\
   0 & v^{(+)}_\ell & 0 \\
\end{array} 
\right), 
\label{Eq:VEV_Sextet}
\end{eqnarray}
and quarks and leptons will acquire masses via these VEV's, where the orthogonal choice of the VEV's in Eq.(\ref{Eq:VEV_Triplet}) and the vanishing VEV of $\langle 0 \vert S^{(+)11} \vert 0\rangle$ in Eq.(\ref{Eq:VEV_Sextet}) are to be ensured by appropriate Higgs interactions.

The pure $SU(3)_L$-anomaly is cancelled in a vectorial manner.  The anomalies from triplets of the three families of leptons and of the three colors of the third family of quarks are cancelled by those from antitriplets of three colors of the first and second families of quarks \cite{SU3U1}.   Other anomalies including $U(1)_N$ are also cancelled. The scale of the 331 model is set by $v_\chi$, yielding $SU(3)_L\times U(1)_N\rightarrow SU(2)_L\times U(1)_Y$ and we expect the magnitude of $v_\chi$ is of order TeV.

\subsection{Interactions}
To realize phenomenologically viable interactions, we further impose a $Z_4$ parity on the model, where $Z_4$=+ for $\psi^1$, $\eta$ and $S^{(+)}$;=$-$ for $\psi^{2,3}$ and $S^{(0)}$;=$i$ for $\rho$ and $\chi$ and similarly for quarks.  Hereafter, we restrict ourselves to the lepton sector. The Yukawa interactions for leptons are, then, controlled by the following lagrangian:
\begin{eqnarray}
&&-{\mathcal{L}}_Y  = \sum_{i=2,3}
f_{1i}{\overline {\left( \psi^1 \right)^c}}\psi^iS^{(0)}
+\sum_{i,j=2,3}
\frac{1}{2}f_{ij}{\overline {\left( \psi^i \right)^c}}\psi^jS^{(+)} 
+ {\rm (h.c.)},
\label{Eq:Yukawa}
\end{eqnarray}
where $f$'s denote the Yukawa couplings with $f_{ij} = f_{ji}$. The possible term of the $\psi\psi\eta$-type is forbidden by $Z_4$. The Higgs interactions are described by self-Hermitian terms composed of $\phi_\alpha\phi^{c \beta}$ ($\phi$=$\eta$, $\rho$, $\chi$) and $S^{(0,+)\alpha\beta}S^{(0,+)c}_{\alpha^\prime\beta^\prime}$, which include the potential term of $V_{\eta\rho\chi}$:
\begin{eqnarray}\label{Eq:Orthgonal}
&V_{\eta\rho\chi}=\lambda_{\eta\rho}\vert \eta \times \rho \vert^2
+\lambda_{\rho\chi}\vert \rho \times \chi \vert^2
+\lambda_{\chi\eta}\vert \chi \times \eta \vert^2,
\end{eqnarray}
where $\lambda$'s represent coupling constants and $(a \times b)^\alpha \equiv \epsilon^{\alpha\beta\gamma}a_\beta b_\gamma$ and by the non-self-Hermitian terms in
\begin{eqnarray}\label{Eq:Conserved}
&V_0 =
\mu_0\rho S^{(0)}\chi 
+\lambda_1\eta\eta S^{(0)c}S^{(0)c}
+\lambda_2\eta S^{(0)}\rho^c\chi^c
+ {\rm (h.c.)},
\end{eqnarray}
where $\mu_0$ denotes a mass scale and $\lambda_{1,2}$ stand for coupling constants.  There is a leptonic number associated with $\psi$, which can be taken to be $-2$ for $S^{(0,+)}$ and 0 for all others.  Since the $L_\ell$ conservation is broken both spontaneously by $S^{(0,+)}$ and explicitly by $V_0$, there is no harmful Nambu-Goldstone boson.

Although the lepton number cannot be globally assigned to $SU(3)_L$-multipulets, to argue Majorana masses for neutrinos, it is useful to define the lepton number $L$ to classify interactions \cite{Lnumbers} in the similar way used for the electric charge defined in the standard $SU(2)_L\times U(1)_Y$ model.  Following the classification lately done in Ref.\cite{Lnumber}, the $L$-number can be introduced in the form of
\begin{eqnarray}\label{Eq:ModifiedLnumber}
&L = 4\lambda^8/2\sqrt{3}+{\mathcal{L}},
\end{eqnarray}
where ${\mathcal{L}}$ = 2 for $J^{1,2}_R$; = 4/3 for $\chi$; = 2/3 for $Q^{1,2}_L$; =1/3 for $\psi^{1,2,3}$; = 0 for $u^{1,2,3}_R$ and $d^{1,2,3}_R$; = $-$2/3 for $Q^3_L$, $\eta$, $\rho$ and $S^{(0,+)}$; = $-$2 for $J^3_R$, which yields $L$ = (1, 1, $-1$) for $\psi^{1,2,3}$ as usual.  The assignment for leptons and Higgs scalars is shown in TABLE \ref{Tab:Lnumber0}.  Among our interactions constrained by $L^\prime$, only $\eta S^{(0)}\rho^c\chi^c$ breaks the $L$ conservation; therefore, it is this interaction that gives the nonvanishing $v^{(0)}_\nu$.  A few comments are in order to support our choice of VEV's of $v_{\eta,\rho,\chi}$ and $v^{(0)}_{\nu,\ell}$:
\begin{enumerate}
\item The nonvanishing VEV of $\langle 0 \vert S^{(0)11} \vert 0\rangle$ is triggered by the $L$-violating $\eta S^{(0)}\rho^c\chi^c$ for $\lambda_2 < 0$ since $\langle 0\vert \eta_1 \vert 0\rangle\neq 0$, $\langle 0\vert \rho_2 \vert 0\rangle\neq 0$ and $\langle 0\vert \chi_3 \vert 0\rangle\neq 0$;
\item The nonvanishing VEV of $\langle 0 \vert S^{(0)23} \vert 0\rangle$ is triggered by the $L$-conserving $\rho S^{(0)}\chi$ for $\mu_0 < 0$ since $\langle 0\vert \rho_2 \vert 0\rangle\neq 0$ and $\langle 0\vert \chi_3 \vert 0\rangle\neq 0$ and by the $L$-conserving $\eta\eta S^{(0)c}S^{(0)c}$ for $\lambda_1 < 0$ since $\langle 0\vert \eta_1 \vert 0\rangle\neq 0$;
\item The orthogonal choice of VEV's of $\eta$, $\rho$ and $\chi$ as in Eq.(\ref{Eq:VEV_Triplet}) is supported by $V_{\eta\rho\chi}$ if all $\lambda$'s are negative.  It is because $V_{\eta\rho\chi}$ gets lowered if $\eta$, $\rho$ and $\chi$ develop VEV's.  So, one can choose VEV's such that $\langle 0\vert \eta_1 \vert 0\rangle$ $\neq$ 0, $\langle 0\vert \rho_2 \vert 0\rangle$ $\neq$ 0 and $\langle 0\vert \chi_3 \vert 0\rangle$ $\neq$ 0.
\end{enumerate}
Thus, these VEV's are dynamically aligned by our Higgs potential. On the other hand, we cannot argue whether $\langle 0 \vert S^{(+)11,23} \vert 0\rangle$ acquire VEV's or not since there are no triggers such as an $L$-violating $\eta S^{(+)}\eta$ and an $L$-conserving $\rho S^{(+)}\chi$, which can, respectively, align $\langle 0 \vert S^{(+)11} \vert 0\rangle$ and $\langle 0 \vert S^{(+)23} \vert 0\rangle$ to be nonvanishing.  However, so far these interactions are forbidden by the $L^\prime$ conservation. To invoke such dynamical alignment, thus, requires $L^\prime$-breaking interactions.

To introduce $L^\prime$-breaking effects calls for careful consideration because of the specific alignment of VEV's of $S^{(+)}$, {\it i.e.} $\langle 0 \vert S^{(+)11} \vert 0\rangle=0$ and $\langle 0 \vert S^{(+)23} \vert 0\rangle\neq 0$.  Its breaking readily induces dangerous interactions that creates $\langle 0 \vert S^{(+)11} \vert 0\rangle$ such as tadpole interactions for $S^{(+)11}$. Owing to the vacuum alignment of $\langle 0 \vert S^{(+)11} \vert 0\rangle=0$, $L^\prime$-breaking interactions should conserve $L$. There is one candidate that can keep this specific alignment, which is an $L$-conserving $\eta\eta S^{(+)c}S^{(+)c}$: 
\begin{eqnarray}\label{Eq:Broken}
&V_b =\lambda_b\eta\eta S^{(+)c}S^{(+)c} +{\rm (h.c)},
\end{eqnarray}
where $\lambda_b$ represents a breaking of the $L^\prime$ conservation.  Since $U(1)_{L^\prime}$ is broken by the unit of $\vert Q_{L^\prime}\vert$=4, there is still a residual conservation due to a $Z_4$ symmetry of $\exp(i\pi Q_{L^\prime}/2 )$ to be denoted by $Z_{L^\prime}$, which is shown in TABLE \ref{Tab:Lnumber0} together with $L^\prime$ and $Z_4$.  This $Z_{L^\prime}$ symmetry can be used to constrain possible forms of $L^\prime$-breaking interactions. The mass terms for $\mu$ and $\tau$ are created by $\langle 0 \vert S^{(+)23} \vert 0\rangle\neq 0$, which is now supported by $V_b$ for $\lambda_b < 0$.  Therefore, $V_b$ serves as a trigger for $\langle 0 \vert S^{(+)23} \vert 0\rangle\neq 0$.

The VEV of $\langle 0 \vert S^{(+)11} \vert 0\rangle$ would be created by the following $L$-violating interactions of 
\begin{itemize}
\item Tr($S^{(0)\dagger} S^{(+)}$) since $\langle 0 \vert S^{(0)11} \vert 0\rangle\neq 0$,
\item $\eta S^{(+)}\eta$ since $\langle 0 \vert \eta_1 \vert 0\rangle\neq 0$,
\item $\eta S^{(+)}\rho^c\chi^c$ since $\langle 0\vert \eta_1 \vert 0\rangle\neq 0$, $\langle 0\vert \rho_2 \vert 0\rangle\neq 0$ and $\langle 0\vert \chi_3 \vert 0\rangle\neq 0$,
\item $\rho\chi S^{(0)c}S^{(+)c}$ and  $\rho\chi S^{(+)c}S^{(+)c}$ since $\langle 0\vert \rho_2 \vert 0\rangle\neq 0$, $\langle 0\vert \chi_3 \vert 0\rangle\neq 0$ and $\langle 0 \vert S^{(0,+)23} \vert 0\rangle\neq 0$ and
\item Tr($S^{(0)}S^{(0)}S^{(+)}$), Tr($S^{(0)}S^{(+)}S^{(+)}$) and Tr($S^{(+)}S^{(+)}S^{(+)}$) since $\langle 0 \vert S^{(0,+)23} \vert 0\rangle\neq 0$.
\end{itemize}
Fortunately, all of these interactions are forbidden by the interplay between $Z_4$ and $Z_{L^\prime}$. Tabulated in TABLE \ref{Tab:Lnumber} are $Z_4$ and $Z_{L^\prime}$ for non-self-Hermitian Higgs interactions allowed by $SU(3)_L\times U(1)_N$, where the reader can recognize how these dangerous interactions are excluded.  At this moment, we reach the plausible situation, where the alignment of
\begin{eqnarray}\label{Eq:NeutralSplus}
&& \langle 0 \vert S^{(+)11} \vert 0\rangle=0, \quad \langle 0 \vert S^{(+)23} \vert 0\rangle \neq 0
\end{eqnarray}
is dynamically supported.

\section{Mass Matrix}
The resulting mass matrices for charged leptons and neutrinos take the form of
\begin{eqnarray}\label{Eq:MassMatrix}
&M_\ell=\left( \begin{array}{ccc}
  0, &  \delta m^\ell_{12}&   \delta m^\ell_{13}\\
  \delta m^\ell_{12}&  m^\ell_{22}&  m^\ell_{23}\\
  \delta m^\ell_{13}&  m^\ell_{23}&  m^\ell_{33}\\
\end{array} \right), \quad
M_\nu=\left( \begin{array}{ccc}
  \delta m^{rad}_{11}&  \delta m^\nu_{12}&   \delta m^\nu_{13}\\
  \delta m^\nu_{12}&  \delta m^{rad}_{22}&  \delta m^{rad}_{23}\\
  \delta m^\nu_{13}&  \delta m^{rad}_{23}&  \delta m^{rad}_{33}\\
\end{array} \right),
\end{eqnarray}
where $m^{\ell, \nu}_{ij}$ are expressed in terms of VEV's:
\begin{eqnarray}\label{Eq:FermionMass}
& \delta m^\ell_{1i}=f_{1i}v^{(0)}_\ell, \quad
\delta m^\nu_{1i}=f_{1i}v^{(0)}_\nu, \quad 
m^\ell_{ij}=f_{ij}v^{(+)}_\ell,
\end{eqnarray}
leading to the proportionality of
\begin{eqnarray}\label{Eq:NuPropto}
&  \delta m^\nu_{1i} \propto \delta m^\ell_{1i},
\end{eqnarray}
and $\delta m^{rad}_{ij}$ stand for one-loop radiative masses to be estimated in the next subsection.  We simply assume $\delta m^\ell_{1i} \ll m^\ell_{ij}$ ($i,j$=2,3) so as to reproduce the hierarchy of $m_e \ll m_{\mu,\tau}$ by the mechanism discussed in Ref.\cite{Hierarchical}. 

Since the magnitude of neutrino masses $\delta m^\nu_{1i}$ is controlled by $v^{(0)}_\nu$, let us explain how naturally small $v^{(0)}_\nu$ is obtained.  An estimate of
\begin{eqnarray}\label{Eq:VEV_nu}
&& v_\nu^{(0)} \sim -\lambda_2 v_\eta v_\rho/v_\chi
\end{eqnarray}
can be found in the analyses of Ref.\cite{NuVEV} applied to our Higgs potential.  It calls for $\lambda_2 < 0$ as has been expected in the vacuum alignment.  The smallness of $v^{(0)}_\nu$ is, then, ascribed to either $\vert\lambda_2\vert \ll 1$ \cite{Lnumber} or $v_\chi\gg v_{\eta,\rho}$ \cite{HugeVEVE}, the latter of which is known as the type II seesaw mechanism \cite{Type2Seesaw}, where we use the mechanism due to $\vert\lambda_2\vert \ll 1$ by simply expecting that $v_\chi$ is of order TeV. One will find that $(-\lambda_2) \sim$a few$\times 10^{-7}$ provides neutrino masses about $5.5\times 10^{-2}$ eV.  This is in fact a fine-tuning of $\lambda_2$ to reproduce neutrino masses so as to meet the value of $5.5\times 10^{-2}$.  However, we have already known this tiny magnitude, which is give by $m_e/m_t$.  The relative smallness of $m_e$ is linked to the underlying chiral symmetry recovered at the limit of $m_e$ $\rightarrow$ 0.  This feature dictates the so-called $\ddash$naturalness" \cite{tHooft}.  Similarly, since the $L$ conservation is respected in the limit of $\lambda_2  \rightarrow 0$, the process of taking $\vert\lambda_2\vert \ll 1$ is also regarded to be natural although both cases require other new physics to explain why such tiny values are realized. 

Our program to discuss patterns of neutrino oscillations is based on the use of the proportionality of $\delta m^\nu_{1i}$ to $\delta m^\ell_{1i}$ indicated by Eq.(\ref{Eq:NuPropto}).  Since to generate the electron mass and to reproduce $\Delta m^2_{atm}$ will, respectively, determine $\delta m^\ell_{1i}$ and $\delta m^\nu_{1i}$, which yield the estimation of $v^{(0)}_\nu/v^{(0)}_\ell$.  The magnitude of $v^{(0)}_\nu$ derived in Eq.(\ref{Eq:VEV_nu}) now determines that of $\lambda_2$ once appropriate values of VEV's are given and, then, $\delta m^{rad}_{11,ij}$ ($i,j$=2,3) to be defined in Eqs.(\ref{Eq:Neutrino_ee}) and (\ref{Eq:Neutrino_mutau}) are calculable for this estimated $\lambda_2$.  In the end, solar neutrino oscillations controlled by $\delta m^{rad}_{11,ij}$ are found to be consistent with the LMA solution.

\subsection{Charged Lepton Masses}
We first discuss the diagonalization of $M_\ell$.  The following unitary matrix $U_\ell$ is obtained under the approximation of $\delta m^\ell_{1i} \ll m^\ell_{ij}$ ($i,j$=2,3) and transforms $M_\ell$ into $M^{diag}_\ell$=$U^\dagger_\ell M_\ell U_\ell$=diag.($-m_e$, $m_\mu$, $m_\tau$):
\begin{eqnarray}\label{Eq:UnitaryMatrix}
&U_\ell=\left( \begin{array}{ccc}
1-\left( \delta U^2_\mu+\delta U^2_\tau\right)/2, 
&
\delta U_\mu
&
\delta U_\tau
\\
-\left(c_\alpha \delta U_\mu + s_\alpha \delta U_\tau \right) 
&
c_\alpha\left( 1-\delta U^2_\mu/2\right)-s_\alpha\delta^2V_{\tau\mu} 
&
s_\alpha\left( 1-\delta U^2_\tau/2\right)+c_\alpha\delta^2V_{\mu\tau}
\\
s_\alpha \delta U_\mu - c_\alpha \delta U_\tau
&
-s_\alpha\left( 1-\delta U^2_\mu/2\right)-c_\alpha\delta^2V_{\tau\mu}
&
c_\alpha\left( 1-\delta U^2_\tau/2\right)-s_\alpha\delta^2V_{\mu\tau}
\\
\end{array} \right),
\end{eqnarray}
up to the second order of $\delta U_{\mu.\tau}$ with
\begin{eqnarray}\label{Eq:UnitaryEntries}
&& \delta U_\mu=\left( 
\left( c^2_\alpha-s^2_\alpha\right) \delta m^\ell_{12}-2c_\alpha s_\alpha \delta m^\ell_{13}\right)/\lambda_-,
\quad
\delta U_\tau=\left( 
2c_\alpha s_\alpha \delta m^\ell_{12} + \left( c^2_\alpha-s^2_\alpha\right) \delta m^\ell_{13}\right)/\lambda_+,
\nonumber \\ 
&& 
\delta^2V_{ij}=\lambda_i\delta U_i\delta U_j/(\lambda_+-\lambda_-), \quad
\lambda_\pm =
\left(
m^\ell_{22} + m^\ell_{33} \pm \sqrt{(m^\ell_{22} - m^\ell_{33})^2 + 4 m^{\ell~2}_{23}}
\right)/2,
\nonumber \\ 
&& 
c_\alpha \equiv \cos\alpha=\sqrt{\left( m^\ell_{33}-\lambda_- \right)/(\lambda_+ - \lambda_-)}, 
\quad 
s_\alpha \equiv \sin\alpha=\sqrt{\left( \lambda_+-m^\ell_{33} \right)/(\lambda_+ - \lambda_-)}, 
\end{eqnarray}
where $i,j$=$\mu,\tau$ for $\delta^2V_{ij}$, $\lambda_\mu=\lambda_-$ and $\lambda_\tau=\lambda_+$.  The masses of $e$, $\mu$ and $\tau$ are given by
\begin{eqnarray}\label{Eq:ChagedLeptonMasses1}
m_e &=& \Delta m_{12}+\Delta m_{13},\quad m_\mu=\lambda_- + \Delta m_{12}
,\quad m_\tau=\lambda_+ + \Delta m_{13},
\end{eqnarray}
where
\begin{eqnarray}\label{Eq:ChagedLeptonMasses2}
&& \Delta m_{12}=\left( 
\left( c^2_\alpha-s^2_\alpha\right) \delta m^\ell_{12}-2c_\alpha s_\alpha \delta m^\ell_{13}
\right)^2/\lambda_-, 
\quad
\Delta m_{13}=\left( 
2c_\alpha s_\alpha \delta m^\ell_{12} + \left( c^2_\alpha-s^2_\alpha\right) \delta m^\ell_{13}
\right)^2/\lambda_+.
\end{eqnarray}

To be consistent with the observed mass pattern of the charged lepton mass matrix, we should present a mechanism generating plausible mass textures such as the hierarchical pattern \cite{Hierarchical} and the democratic pattern \cite{Democratic}.  However, in the present context of the model, we are not aimming at clarifying all the physical aspect of the lepton sector but, instead, at clarifying the correlation of neutrino mass matrix with charged lepton mass matrix.  Here, we simply parameterize $m^\ell_{ij}$ by adopting two distinct ways to realize $m_\mu \ll m_\tau$: hierarchical and democratic mass textures and discuss their influences in neutrino mixings.  The resulting parameterization is given by
\begin{enumerate}
\item the hierarchical pattern of $m^\ell_{22,23} \ll m^\ell_{33}$ and
\begin{eqnarray}\label{Eq:Masses23_1}
&& m_e \approx \delta m^{\ell~2}_{12}/m_\mu + \delta m^{\ell~2}_{13}/m_\tau,
\quad
m_\mu \approx m^\ell_{22} + \left( m^{\ell~2}_{23}/m_\tau \right),
\quad 
m_\tau \approx m^\ell_{33} - \left( m^{\ell~2}_{23}/m_\tau \right),
\nonumber \\
&& \delta U_\mu \approx \delta m^\ell_{12}/m_\mu, 
\quad
\delta U_\tau \approx \delta m^\ell_{13}/m_\tau, 
\end{eqnarray}
leading to $\cos\alpha \approx 1$ with $\sin\alpha \approx m^\ell_{23}/m^\ell_{33}$ ($\ll 1$) and
\item the democratic pattern of $m^\ell_{22} \approx  m^\ell_{23} \approx m^\ell_{33}$ and
\begin{eqnarray}\label{Eq:Masses23_2}
&& m_e \approx \delta m^{\ell~2}_{13}/m_\mu +\delta  m^{\ell~2}_{12}/m_\tau,
\quad
m_\mu \approx \Delta m,
\quad 
m_\tau \approx 2m_0 - \Delta m,
\nonumber \\
&& \delta U_\mu \approx -\delta m^\ell_{13}/m_\mu, 
\quad
\delta U_\tau \approx \delta m^\ell_{12}/m_\tau, 
\end{eqnarray}
where $m_0$ and $\Delta m$ specify $m^\ell_{22}=m^\ell_{33}=m_0$ and $m^\ell_{23}=m_0 - \Delta m$ ($m_0 \gg \Delta m$), leading to $\cos\alpha = \sin\alpha = 1/\sqrt{2}$.
\end{enumerate}
Therefore, we have found the estimate of $\delta m^\ell_{12.13}$ that reproduces $m_e$.

\subsection{Radiative Masses}
Neutrinos acquire radiative masses of $\delta m^{rad}_{11,ij}$ ($i,j$=2,3) generated by one-loop interactions corresponding to diagrams depicted in FIG.\ref{Fig:loopDiagrams_ee} and FIG.\ref{Fig:loopDiagrams_mutau}.  It is convenient to shift the basis for charged-leptons from the original one for $M_\ell$ to the intermediate one, where $\mu$ and $\tau$ have diagonal masses of $\lambda_-$ and $\lambda_+$, respectively.  The corrections due to $\delta m^\ell_{12,13}$ are neglected because of $\delta m^\ell_{12,13}\ll m_{\mu,\tau}$. In this basis, the couplings of $S^{(0)}$ to the lepton triplets are described by $F_{1i}\psi^1\Psi^i$ ($i$=2,3) with ($\Psi^2$, $\Psi^3$)=($c_\alpha \psi^2 - s_\alpha\psi^3$, $s_\alpha \psi^2 + c_\alpha\psi^3$) and $F_{1i}$ are the rotated  $f_{1i}$-couplings similarly defined as ($F_{12}$, $F_{13}$)=$(c_\alpha f_{12}-s_\alpha f_{13}$, $s_\alpha f_{12}+c_\alpha f_{13}$), which appear in the figures.

The result of the calculation of $\delta m^{rad}_{11,ij}$ ($i,j$=2,3) is given by
\begin{eqnarray}
\delta m^{rad}_{11}&=&
2\lambda_2
\mu_0 
\sum_{i=2,3} F^2_{1i} m_{\ell^i}
\left[
I(m^2_{\ell^i}, m^2_{s^{(0)+}}, m^2_{s^{(0)-}}, m^2_{\rho^+ })v^2_\chi
+I(m^2_{\ell^i}, m^2_{s^{(0)+}}, m^2_{s^{(0)-}}, m^2_{\chi^-} )v^2_\rho
\right]v_\eta
\nonumber \\
&&+
2\lambda_2\lambda_1 v^{(0)}_\ell 
F^2_{1i} m_{\ell^i}
\left[
I(m^2_{\ell^i}, m^2_{s^{(0)+}}, m^2_{s^{(0)-}}, m^2_{\eta^+} )
+I(m^2_{\ell^i}, m^2_{s^{(0)-}}, m^2_{s^{(0)+}}, m^2_{\eta^- })
\right] v_\eta v_\rho v_\chi,
\label{Eq:Neutrino_ee} \\
\delta m^{rad}_{ij}&=&
2\lambda_2\lambda_b  v^{(0)}_\ell 
F_{1i}F_{1j} m_{\ell^i}
\left[
I(m^2_{\ell^i}, m^2_{s^{(+)+}}, m^2_{s^{(0)-}}, m^2_{\eta^+} )
+I(m^2_{\ell^i}, m^2_{s^{(+)-}}, m^2_{s^{(0)+}}, m^2_{\eta^- })
\right] v_\eta v_\rho v_\chi,
\label{Eq:Neutrino_mutau}
\end{eqnarray}
for $i,j$=2,3 and $m_e$=0 in this basis, with $m_{\ell^2}=\lambda_-\approx m_\mu$ and $m_{\ell^3}=\lambda_+\approx m_\tau$ as in Eq.(\ref{Eq:UnitaryEntries}), where
\begin{eqnarray}
&& I(a, b, c, d)=\frac{1}{16\pi^2}\frac{ J(a, b, d)-J(a, c, d)}{ b-c},
\label{Eq:Integral} \\
&& J(a,b,c)=\frac{a\ln a}{\left( a-b\right)\left( a-c\right)}
+\frac{b\ln b}{\left( b-a\right)\left( b-c\right)}
+\frac{c\ln c}{\left( c-b\right)\left( c-a\right)}.
\label{Eq:Integral_oneloop}
\end{eqnarray}
The Majorana mass $\delta m^{rad}_{11}$ for $\nu_e$ is induced by effective couplings corresponding to FIG.\ref{Fig:loopDiagrams_ee}:
\begin{eqnarray}\label{Eq:InducedMajorana_ee}
&&(\eta^\dagger\psi^1)\epsilon^{\alpha\beta\gamma}\psi^1_\alpha\phi_\beta (\phi S^{(+)})^c_\gamma,
\end{eqnarray}
for $\phi$=$\rho$ or $\chi$, and 
\begin{eqnarray}\label{Eq:InducedMajorana_ee2}
&&
\epsilon^{\alpha\beta\gamma}\epsilon^{\alpha^\prime\beta^\prime\gamma^\prime}\psi^1_\alpha\psi^1_{\alpha^\prime} (S^{(+)})^c_{\beta\beta^\prime}(S^{(0)})^c_{\gamma\gamma^\prime}
\left( \epsilon^{\alpha\beta\gamma}\eta_\alpha\rho_\beta\chi_\gamma\right)
\end{eqnarray}
with similar terms scrambled by possible permutations of the fields.  Both couplings conserve $L^\prime$. On the other hand, the same effective coupling as Eq.(\ref{Eq:InducedMajorana_ee2}) but with the replacement of $\lambda_2\rightarrow \lambda_b$, thus of the $L^\prime$-breaking type, yields $\delta m^{rad}_{ij}$ ($i,j$=2,3) for $\nu_{\mu,\tau}-\nu_{\mu,\tau}$, which correspond to FIG.\ref{Fig:loopDiagrams_mutau}.

\subsection{Neutrino Masses}
Now, let us turn to examining the neutrino sector. The neutrino mass matrix is also 
transformed into $M^{weak}_\nu$ by $U_\ell$ of Eq.(\ref{Eq:UnitaryMatrix}) to maintain 
diagonal weak currents and is calculated, up to the second order of $\delta U_{\mu,\tau}$, to be:
\begin{eqnarray}\label{Eq:NuMassesWeak}
&& M^{weak}_\nu=U^\dagger_\ell M_\nu U_\ell \approx 
\left( \begin{array}{ccc}
  \delta M_{11}, &  M_{12}& M_{13}\\
  M_{12}&  \delta M_{22}&  \delta M_{23}\\
  M_{13}&  \delta M_{23}&  \delta M_{33}\\
\end{array} \right),
\end{eqnarray}
where
\begin{eqnarray}
M_{12}&=&N_{12} \left( 1-2\delta U^2_\mu-\delta U^2_\tau/2\right)
-
N_{13}
\left(
\delta^2 V_{\tau\mu}
+
\delta U_\mu\delta U_\tau
\right),
\nonumber \\ 
M_{13}&=&N_{13} \left( 1-2\delta U^2_\tau-\delta U^2_\mu/2\right)
+
N_{12}
\left( 
\delta^2 V_{\mu\tau}
-
\delta U_\mu\delta U_\tau
\right)
,
\nonumber \\ 
\delta M_{11}&=&\delta m^{rad}_{11}-2\left( N_{12} \delta U_\mu+N_{13} \delta U_\tau\right),
\nonumber \\ 
\delta M_{22}&=&2N_{12} \delta U_\mu 
+ c^2_\alpha \delta m^{rad}_{22}
- 2c_\alpha s_\alpha \delta m^{rad}_{23}
+ s^2_\alpha \delta m^{rad}_{33},
\nonumber \\ 
\delta M_{23}&=&N_{13} \delta U_\mu + N_{12}\delta U_\tau
+ c_\alpha s_\alpha\left( \delta m^{rad}_{22}-\delta m^{rad}_{33}\right)
+
\left( c^2_\alpha - s^2_\alpha \right) \delta m^{rad}_{23},
\nonumber \\ 
\delta M_{33}&=&2N_{13} \delta U_\tau
+ s^2_\alpha \delta m^{rad}_{22}
+ 2c_\alpha s_\alpha \delta m^{rad}_{23}
+ c^2_\alpha \delta m^{rad}_{33}
\label{Eq:MassesNuEntry}
\end{eqnarray}
with 
\begin{eqnarray}\label{Eq:C_12C_13}
&&N_{12}=c_\alpha \delta m^\nu_{12} - s_\alpha \delta m^\nu_{13},
\quad 
N_{13}=c_\alpha \delta m^\nu_{13} + s_\alpha \delta m^\nu_{12}.  
\end{eqnarray}
It is obvious to see that $M^{weak}_\nu$ satisfies Tr($M^{weak}_\nu$)=Tr($M_\nu$)=$\delta m^{rad}_{11}+\delta m^{rad}_{22}+\delta m^{rad}_{33}$ and Tr($M^{weak~T}_\nu M^{weak}_\nu$)= Tr($M^T_\nu M_\nu$)=2($\delta m^{\nu~2}_{12}+\delta m^{\nu~2}_{13})$.  Since $\vert M_{12,13}\vert \gg \vert \delta M_{ij}\vert$, the diagonal neutrino masses given by  ($m_{\nu 1}$, $-m_{\nu 2}$, $m_{\nu 3}$) are calculated to be:
\begin{eqnarray}\label{Eq:NeutrinoMasses}
&& m_{\nu 1}=m_\nu + m^{(1)}_{\nu 1} + m^{(2)}_\nu,
\quad
m_{\nu 2}=m_\nu - m^{(1)}_{\nu 1} + m^{(2)}_\nu,
\quad
m_{\nu 3}=m^{(1)}_{\nu 3},
\end{eqnarray}
up to the second order of $\delta M_{ij}$ equivalent to the second order of $\delta U_{\mu,\tau}$, where
\begin{eqnarray}\label{Eq:NeutrinoMassParam}
&& m_\nu=\sqrt{M^2_{12} + M^2_{13}},
\quad
m^{(1)}_{\nu 1}=\left(\delta M_{11}+\cos^2\vartheta_\nu \delta M_{22}+2\cos\vartheta_\nu\sin\vartheta_\nu \delta M_{23}+\sin^2\vartheta_\nu \delta M_{33}\right)/2,
\nonumber \\ 
&&m^{(1)}_{\nu 3}=\sin^2\vartheta_\nu\delta M_{22}-2\cos\vartheta_\nu\sin\vartheta_\nu \delta M_{23}+\cos^2\vartheta_\nu \delta M_{33},
\quad
m^{(2)}_\nu=\frac{M^{\prime 2}_{12}}{2m_\nu}+
\frac{M^{\prime 2}_{13}}{m_\nu}
\end{eqnarray}
with $\cos\vartheta_\nu=M_{12}/m_\nu$ and $\sin\vartheta_\nu=M_{13}/m_\nu$ as the mixing angle for atmospheric neutrinos and
\begin{eqnarray}\label{Eq:NeutrinoMassesHigherOrderElemnt}
&&M^\prime_{12}=\left(-\delta M_{11}+\cos^2\vartheta_\nu \delta M_{22}+2\cos\vartheta_\nu\sin\vartheta_\nu \delta M_{23}+\sin^2\vartheta_\nu \delta M_{33}\right)/2,
\nonumber \\
&&M^\prime_{13}=\left( \left( \cos^2\vartheta_\nu -\sin^2\vartheta_\nu\right) \delta M_{23}+\cos\vartheta_\nu\sin\vartheta_\nu \left( \delta M_{33}-\delta M_{22}\right)\right)/\sqrt{2}.
\end{eqnarray}
The masses of $m^{(1)}_{\nu 1}$ and $M^\prime_{12,13}$ can be further converted into: 
\begin{eqnarray}
m^{(1)}_{\nu 1} &=& 
\frac{\delta m^{rad}_{11}+\delta m^{rad}_{22}}{2}
+
\frac{
\left( 
 M_{13}\delta U_\mu -M_{12} \delta U_\tau
\right) 
\left(
M_{12}N_{13}-M_{13}N_{12}
\right)
}
{
m^2_\nu
},
\label{Eq:EachMassTerms1} \\ 
M^\prime_{12}&=&
N_{12}\delta U_\mu+N_{13}\delta U_\tau
+\frac{
\left(
M_{12}\delta U_\mu+M_{13}\delta U_\tau
\right)
\left(
M_{12}N_{12}+M_{13}N_{13}
\right)
}
{m^2_\nu},
\label{Eq:EachMassTerms2} \\ 
M^\prime_{13}
&=&
\frac{1}{\sqrt{2}}
\left[
N_{13}\delta U_\mu-N_{12}\delta U_\tau
-2
\frac{
\left(
M_{13}\delta U_\mu-M_{12}\delta U_\tau
\right)
\left(
M_{12}N_{12}+M_{13}N_{13}
\right)
}
{m^2_\nu}
\right],
\label{Eq:EachMassTerms3}
\end{eqnarray}
where other irrelevant terms involving $\delta m^{rad}_{ij}$ are omitted.  The mass of $m^{(1)}_{\nu 3}$ is obtained by noticing the relation of Tr($M^{weak}_\nu$)=$\delta m^{rad}_{11}+\delta m^{rad}_{22}+\delta m^{rad}_{33}$ that yields
\begin{eqnarray}
2m^{(1)}_{\nu 1}+m^{(1)}_{\nu 3}&=&\delta m^{rad}_{11}+\delta m^{rad}_{22}+\delta m^{rad}_{33},
\label{Eq:SumRules}
\end{eqnarray}
from which we find that
\begin{eqnarray}
m^{(1)}_{\nu 3}=\delta m^{rad}_{33}
-2\frac{
\left( 
 M_{13}\delta U_\mu -M_{12} \delta U_\tau
\right) 
\left(
M_{12}N_{13}-M_{13}N_{12}
\right)
}
{
m^2_\nu
}.
\label{Eq:ThirdMass}
\end{eqnarray}

To enhance the bimaximal structure in $M^{weak}_\nu$ is achieved by setting $\vert M_{12}\vert \approx \vert M_{13}\vert$, where $M_{12,13}$ can be approximated to be:
\begin{eqnarray}\label{Eq:M_1i}
&& M_{12} \approx N_{12} = c_\alpha \delta m^\nu_{12} - s_\alpha \delta m^\nu_{13},
\quad
M_{13} \approx N_{13} = c_\alpha \delta m^\nu_{13} + s_\alpha \delta m^\nu_{12},
\end{eqnarray}
because $\vert\delta U_{\mu,\tau}\vert \ll 1$.  The condition of $\vert M_{12}\vert \approx \vert M_{13}\vert$ thus, requires, for the hierarchical case with $c_\alpha \approx 1$ and $s_\alpha \approx 0$,
\begin{eqnarray}\label{Eq:BimaximalCond1}
&& \delta m^\nu_{12} \approx \delta m^\nu_{13},
\end{eqnarray}
and, for the democratic case with $c_\alpha = s_\alpha = 1/\sqrt{2}$,
\begin{eqnarray}\label{Eq:BimaximalCond2}
&& \delta m^\nu_{12}  \gg  \delta m^\nu_{13}\quad {\rm or} \quad
\delta m^\nu_{12}  \ll \delta m^\nu_{13}.
\end{eqnarray}
The essence of the present research lies in the fact that the bimaximal mixing scheme for neutrinos, in turn, determines how the electron mass is derived.  The bimaximal condition of Eq.(\ref{Eq:BimaximalCond1}) combined with Eq.(\ref{Eq:Masses23_1}) yields
\begin{eqnarray}\label{Eq:BimaximalChargedCond1}
&& \delta m^\ell_{12} \approx \delta m^\ell_{13} \approx \sqrt{m_em_\mu},
\quad
\delta U_\mu \approx \sqrt{m_e/m_\mu},
\quad
\delta U_\tau \approx \sqrt{m_e/m_\tau},
\end{eqnarray}
while Eq.(\ref{Eq:BimaximalCond2}) combined with Eq.(\ref{Eq:Masses23_2}) yields 
\begin{eqnarray}
&& 
\delta m^\ell_{12}  \approx \cos\phi\sqrt{m_em_\tau}, 
\quad
\delta m^\ell_{13}\approx \sin\phi\sqrt{m_em_\mu},
\quad
\delta U_\mu \approx -\sin\phi\sqrt{m_e/m_\mu},
\quad
\delta U_\tau \approx \cos\phi\sqrt{m_e/m_\tau},
\label{Eq:BimaximalChargedCond2}
\end{eqnarray}
and
\begin{eqnarray}
&&
\delta m^\ell_{12}  \approx \varepsilon_\ell\sqrt{m_em_\mu},
\quad
\delta m^\ell_{13}\approx \sqrt{m_em_\mu},
\quad
\delta U_\mu \approx -\sqrt{m_e/m_\mu},
\quad
\delta U_\tau \approx \varepsilon_\ell\sqrt{m_em_\mu}/m_\tau,
\label{Eq:BimaximalChargedCond3}
\end{eqnarray}
where $\tan^2\phi \ll m_\tau/m_\mu$.

\section{Neutrino Oscillations}
The neutrino oscillation parameters of $\Delta m^2_{atm}$ and $\Delta m^2_{\odot}$ are expressed in terms of our neutrino masses as
\begin{eqnarray}\label{Eq:Atm_m2}
&& \Delta m^2_{atm} \equiv m^2_{\nu 2}-m^2_{\nu 3}\approx m^2_\nu \approx M^2_{12} + M^2_{13} \approx \delta m^{\nu~2}_{12}+\delta m^{\nu~2}_{13}, 
\\ \label{Eq:Solar_m2}
&& \Delta m^2_{\odot} \equiv m^2_{\nu 1}-m^2_{\nu 2} \approx 
4m_\nu m^{(1)}_{\nu 1}=2m_\nu \left( \delta m^{rad}_{11}+\delta m^{rad}_{22}+\delta m^{rad}_{33}  - m^{(1)}_{\nu 3}\right),
\end{eqnarray}
leading to
\begin{eqnarray}\label{Eq:RatioSolarOverAtm}
&&
\frac{\Delta m^2_{\odot}}{\Delta m^2_{atm}} 
=
\left(\frac{\Delta m^2_{\odot}}{\Delta m^2_{atm}}\right)_{rad}
+
\left(\frac{\Delta m^2_{\odot}}{\Delta m^2_{atm}}\right)_{rot}
\end{eqnarray}
with
\begin{eqnarray}\label{Eq:RatioSolarRad}
&&\left(\frac{\Delta m^2_{\odot}}{\Delta m^2_{atm}}\right)_{rad}=
\frac{
2\left( \delta m^{rad}_{11}+\delta m^{rad}_{22}\right) 
}
{
m_\nu
},
\quad
\left(\frac{\Delta m^2_{\odot}}{\Delta m^2_{atm}}\right)_{rot}=
4\frac{
\left( 
 M_{13}\delta U_\mu -M_{12} \delta U_\tau
\right) 
\left(
M_{12}N_{13}-M_{13}N_{12}
\right)
}
{
m^3_\nu
}
.
\end{eqnarray}
The ratio $(\Delta m^2_{\odot}/\Delta m^2_{atm})_{rad}$ arises from the radiative effects and $(\Delta m^2_{\odot}/\Delta m^2_{atm})_{rot}$ receives the first order contributions from the rotation due to the diagonalization of $M_{\nu,\ell}$.

We are now in a position to estimating  $\Delta m^2_{\odot}/\Delta m^2_{atm}$ to meet the LMA solution.  Before going to its explicit evaluation, we list our assumptions on numerical values of the relevant mass parameters:
\begin{itemize}
    \item $v_\chi \gg v_{weak}$ (= $( 2{\sqrt 2}G_F)^{-1/2}$=174 GeV) for $SU(3)_L\times U(1)_N \rightarrow SU(2)_L\times U(1)_Y$,
    \item $v_\eta=v_\rho=v^{(0)}_\ell=v^{(+)}_\ell=v_{weak}/2$ as the simplest case to keep $v^2_\eta + v^2_\rho + v^{(0)2}_\ell + v^{(+)2}_\ell=v^2_{weak}$ for weak boson masses,
    \item $m_{\eta^\pm}=m_{\rho^+}=m_{s^{(0)\pm}}=m_{s^{(+)\pm}}=v_{weak}$ and $m_{\chi^+}=\vert\mu_0\vert=v_\chi$,
\end{itemize}
where $v_\chi$, presumably of order TeV, is left unspecified and is to be measured in the unit of 10$v_{weak}$.  

\subsection{Atmospheric Neutrinos}
Let us first reproduce $\Delta m^2_{atm}\sim 3\times 10^{-3}$ eV$^2$, which requires that $\delta m^{\nu~2}_{12}+\delta m^{\nu~2}_{13}\sim 3\times 10^{-3}$ eV$^2$ from Eq.(\ref{Eq:Atm_m2}).  The relation of $\delta m^\nu_{1i}=f_{1i}v^{(0)}_\nu=\delta m^\ell_{1i} v^{(0)}_\nu/v^{(0)}_\ell$ ($i$=2,3) together with Eqs.(\ref{Eq:BimaximalChargedCond1}-\ref{Eq:BimaximalChargedCond3}) for $\delta m^\ell_{1i}$, then, leads to
\begin{eqnarray}\label{Eq:VEV_nuNumeric}
&& 
v_\nu^{(0)} \approx \left(5.5, 1.8/\cos\phi, 7.5\right)\times 10^{-9}v^{(0)}_\ell,
\end{eqnarray}
where three terms to be placed in the parenthesis, respectively, denote the cases with (hierarchical masses with $\delta m^\nu_{12} \approx \delta m^\nu_{13}$, democratic masses with $\delta m^\nu_{12} \gg \delta m^\nu_{13}$, democratic masses with $\delta m^\nu_{12} \ll \delta m^\nu_{13}$). In the following, the same notation is used.  Since $v^{(0)}_\nu$ is expressed in terms of $\lambda_2$ and VEV's as estimated in Eq.(\ref{Eq:VEV_nu}), we find the constraint on $\lambda_2$ given by
\begin{eqnarray}\label{Eq:lambda2_estimate}
&& 
-\lambda_2 \approx \left(1.1, 0.36/\cos\phi, 1.5 \right)
v_\chi/10v_{weak} \times 10^{-7}.
\end{eqnarray}
Therefore, the magnitude of $\lambda_2$ is not arbitrary but fixed to be $\vert\lambda_2\vert ={\mathcal{O}}(10^{-7})$ by VEV's in order to obtain $\Delta m^2_{atm}\sim 3\times 10^{-3}$ eV$^2$.

\subsection{Solar Neutrinos}
Next, we calculate $\Delta m^2_\odot$, which depends on the sum of $\delta m^{rad}_{11}$ and $\delta m^{rad}_{22}$ given by Eqs.(\ref{Eq:Neutrino_ee}) and (\ref{Eq:Neutrino_mutau}) with $i=j=2$.  To find numerical values of $\delta m^{rad}_{11,ij}$, we use the following property of the integral $I$ of Eq.(\ref{Eq:Integral}):
\begin{eqnarray}\label{Eq:ApproximatedIntegral_oneloop1}
&& I(0, a, a, b)
\approx
-1/ab(=-I_1),
\quad
I(0, a, a, a)
\approx
-1/a^2(=-I_2).
\end{eqnarray}
By using the assumptions on mass parameters stated above, we know that the second term in Eq.(\ref{Eq:Neutrino_ee}) is proportional to $I_1$ and all other terms including $\delta m^{rad}_{ij}$ of Eq.(\ref{Eq:Neutrino_mutau}) are proportional to $I_2$. In Eq.(\ref{Eq:Neutrino_ee}), the first term thus dominates the second term because of $I_2\gg I_1$ for $m^2_{\chi^-} \gg m^2_{s^{(0)+}}$ and $v_\chi \gg v_\rho$.  The other terms proportional to $I_2$, the third and fourth terms together with $\delta m^{rad}_{22}$, are more suppressed by the factor of $\lambda_{1,b}v^2_{weak}/ v^2_\chi$ compared with the first term since $\vert\lambda_{1,b}\vert$ are at most of order unity. Therefore, the first term in $\delta m^{rad}_{11}$ gives the most dominant contributions to $\Delta m^2_\odot$ and those from all other terms can be safely neglected.  The radiative mass of $\delta m^{rad}_{11}$ saturated by the mediating $\tau$ lepton contribution turns out to be:
\begin{eqnarray}\label{Eq:Numerics_Rad1}
&& 
\delta m^{rad}_{11} \approx -73\lambda_2 
\left( 1, \cos^2\phi m_\tau/2m_\mu, 1/2\right)
\left( v_\chi/10v_{weak}\right)^3~{\rm eV},
\end{eqnarray}
where, in the democratic cases, the factor of $m_\tau\cos^2\phi/m_\mu$ is due to $\delta m^{\ell~2}_{12} (\propto f^2_{12})=m_em_\tau\cos^2\phi$ in Eq.(\ref{Eq:BimaximalChargedCond2}) and the factor 1/2 arises from $c^2_\alpha = s^2_\alpha = 1/2$.  The rotated couplings of $F_{1i}$ used in these calculations are approximated to be ($f_{1i}$, $f_{12}/\sqrt{2}$, $\Theta f_{13}/\sqrt{2}$) ($i=2,3$) for $\Theta$=$-1$ for $i$=2 and $\Theta$=1 for $i$=3.

By inserting the above result of Eq.(\ref{Eq:Numerics_Rad1}) into $(\Delta m^2_{\odot}/\Delta m^2_{atm})_{rad}$ shown in Eq.(\ref{Eq:RatioSolarRad}) with $m_\nu\sim 5.5\times 10^{-2}$ eV, we reach the following estimate:
\begin{eqnarray}\label{Eq:Numerics_Rad2}
&& 
\left(\Delta m^2_{\odot}/\Delta m^2_{atm}\right)_{rad} \approx 
-2.7\lambda_2 
\left( 1, 8.5\cos^2\phi, 0.5\right)
\left( v_\chi/10v_{weak}\right)^3 \times 10^3,
\end{eqnarray}
where $m_\tau/m_\mu\approx 17$ is used in Eq.(\ref{Eq:Numerics_Rad1}).  This estimate together with Eq.(\ref{Eq:lambda2_estimate}) for $\lambda_2$ yields
\begin{eqnarray}\label{Delta_sol_chi}
&& 
\left(\Delta m^2_{\odot}/\Delta m^2_{atm}\right)_{rad}
\approx \left(2.9, 8.1\cos\phi, 1.6 \right)
\left(v_\chi/10v_{weak} \right)^4 \times 10^{-4},
\end{eqnarray}
where $\cos\phi$ is bounded from below by $\tan^2\phi \ll m_\tau/m_\mu$. By taking
\begin{eqnarray}\label{Delta_sol_LMA}
&& 
v_\chi/10v_{weak}
\approx \left(1.8-3.3, (1.4-2.5)\sqrt[4]{\cos\phi}, 2.0-3.6 \right),
\end{eqnarray}
we obtain $\Delta m^2_{\odot}$=($10^{-5}-10^{-4}$) eV$^2$ relevant to the LMA solution to the solar neutrino problem and
\begin{eqnarray}\label{RadiativeMass_sol_LMA}
&& 
\delta m^{rad}_{11}=\Delta m^2_\odot/2m_\nu \approx 10^{-4} - 10^{-3}~{\rm  eV}.
\end{eqnarray}
Since $v_{weak}$=174 GeV, $v_\chi$ roughly lies in the range of $3-6$ TeV, which is the right order of $v_\chi$ (= ${\mathcal{O}}$(1 TeV)) as anticipated for the scale of $SU(3)_L\times U(1)_N$ $\rightarrow$ $SU(2)_L\times U(1)_Y$.  For $v_\chi= 4$ TeV as a typical value, namely, $v_\chi/10v_{weak}$=2.3, we find that
\begin{eqnarray}\label{Delta_sol_LMA2}
&& 
\left(\Delta m^2_{\odot}/\Delta m^2_{atm}\right)_{rad}
\approx \left(0.8, 2.3\cos\phi,0.6 \right)\times 10^{-2},
\end{eqnarray}
leading to $\Delta m^2_{\odot}\sim (2.4, 6.9\cos\phi, 1.8) \times 10^{-5}$ eV$^2$ for $\Delta m^2_{atm}\sim 3\times10^{-3}$ eV$^2$.

\subsection{Diagonalization Effect}
Finally, we consider effects from $(\Delta m^2_{\odot}/\Delta m^2_{atm})_{rot}$, which may disturb the behavior specific to the LMA solution indicated by $(\Delta m^2_{\odot}/\Delta m^2_{atm})_{rad}$.  Since $(\Delta m^2_{\odot}/\Delta m^2_{atm})_{rot}$ is controlled by $m^{(1)}_{\nu 3}$ as in Eq.(\ref{Eq:RatioSolarRad}), it can be found that the leading terms of ${\mathcal{O}}(\delta U_{\mu,\tau})$ contained in $m^{(1)}_{\nu 3}$ turns out to vanish.  The cancellation can be seen from the insertion of the leading terms of $M_{12,13}\approx N_{12,13}$ into the equation of $M_{12}N_{13}-M_{13}N_{12}$ appearing in the numerator of $m^{(1)}_{\nu 3}$ in Eq.(\ref{Eq:ThirdMass}) that turns out to vanish. The next nonvanishing contributions to $m^{(1)}_{\nu 3}$ are given by the second order terms in $M_{12}N_{13}-M_{13}N_{12}$ that result in the third order contributions to $m^{(1)}_{\nu 3}$, which is beyond our approximation because possible third order terms for $\delta M_{ij}$ ($i,j$=2,3), which have not been included, directly give the third order contributions as can be seen from the definition of $m^{(1)}_{\nu 3}$ in Eq.(\ref{Eq:NeutrinoMassParam}). Therefore, we expect that $(\Delta m^2_{\odot}/\Delta m^2_{atm})_{rot}$ becomes at most ${\mathcal{O}}(\delta U^3_\mu)$, which is about $(m_e/m_\mu)^{3/2}(=3.4\times 10^{-4}$) as shown in Eqs.(\ref{Eq:BimaximalChargedCond1})-(\ref{Eq:BimaximalChargedCond3}), leading to $(\Delta m^2_{\odot}/\Delta m^2_{atm})_{rot} \mapleq 10^{-4}$. This estimate of $(\Delta m^2_{\odot}/\Delta m^2_{atm})_{rot}$ should be compared with Eq.(\ref{Delta_sol_LMA2}) for the radiative case and in fact gives more suppressed contributions.  It is, then, concluded that we can safely neglect $(\Delta m^2_{\odot}/\Delta m^2_{atm})_{rot}$ and that the model can provide the LMA solution to the solar neutrino problem as indicated by $(\Delta m^2_{\odot}/\Delta m^2_{atm})_{rad}$.

Similarly, the third neutrino $\nu_3$ will also acquire masses at most of order $m_\nu \delta U^3_\mu$, which is about $10^{-5}$ eV.  The radiative correction to $m_{\nu 3}$ arising from $\delta m^{rad}_{33}$ is estimated to be about
\begin{eqnarray}\label{mass_nu3}
&& 
\delta m^{rad}_{33} \sim -0.2\lambda_2\lambda_b \left(1, m_\tau\cos^2\phi /2m_\mu, 0.5\right)\left(v_\chi/10v_{weak}\right)~{\rm eV}.
\end{eqnarray}
Since $\lambda_2\sim 10^{-7}$ and $\lambda_b\mapleq 1$, we find that $\delta m^{rad}_{33}\mapleq 0.5\times 10^{-7}$ eV for $v_\chi\sim 4$ TeV.  Thus, altogether we observe that the bound on $m_{\nu 3}$ is given by $m_{\nu 3}\mapleq 10^{-5}$ eV.

The cancellation of the diagonalization effects on the neutrino masses of $m^{(1)}_{\nu 1,\nu 3}$ is a reasonable consequence because the (1,2) and (1,3) entries of $M_\nu$ are exactly proportional to those of $M_\ell$ (see Eqs.(\ref{Eq:LeptonTextureH})-(\ref{Eq:LeptonTextureB2})) so that the rotation effects due to $\delta U_{\mu,\tau}$, which induce $\delta M_{22}=2m_\nu\cos\vartheta_\nu\delta U_\mu$, $\delta M_{23}=m_\nu (\cos\vartheta_\nu\delta U_\mu+\sin\vartheta_\nu\delta U_\tau )$, $\delta M_{33}=2m_\nu\sin\vartheta_\nu\delta U_\tau$, are partially cancelled.  However, the diagonalization effects manifest themselves in the matrix element $U_{e3}$ that connects $\nu_e$ to $\nu_3$. The CHOOZ and PALOVERDE experiments imply that $\vert U_{e3}\vert^2\mapleq 0.015-0.05$ \cite{Chooz}.  For the mass matrix $M^{weak}_\nu$ of Eq.(\ref{Eq:NuMassesWeak}), this element is calculated to be
\begin{eqnarray}
U_{e3} &=& -\frac{1}{m_\nu}\left[
\left( \cos^2\vartheta_\nu-\sin^2\vartheta_\nu\right)\delta M_{23}+\cos\vartheta_\nu\sin\vartheta_\nu\left( \delta M_{33}-\delta M_{22}\right)
\right]
\nonumber \\
&=& -\frac{1}{m_\nu}\left[
\left(
c_\alpha \delta m^\nu_{13} + s_\alpha \delta m^\nu_{12}  
\right)\delta U_\mu
-
\left(
c_\alpha \delta m^\nu_{12} - s_\alpha \delta m^\nu_{13}
\right)\delta U_\tau
\right],
\label{Eq:Ue3}
\end{eqnarray}
leading to $\delta U_\mu/\sqrt{2}\approx (1, -\sin\phi, -1)\sqrt{m_e/2m_\mu}$ with $m_\mu/m_\tau \ll \tan^2\phi \ll m_\tau/m_\mu$ for Eq.(\ref{Eq:BimaximalChargedCond2}) allowing $\vert\delta U_\mu\vert \gg \vert\delta U_\tau\vert$, which includes the typical value of $\tan\phi\sim 1$.  Our model predicts that $\vert U_{e3}\vert^2\approx (1, \sin^2\phi, 1)\times 0.0025$, which are consistent with the current experimental bound.

\section{Summary and Discussions}
We have successfully demonstrated that the minimal 331 model with the global $L_e-L_\mu-L_\tau$ and $Z_4$ symmetry, which involves two (anti-)sextet Higgs scalars, $S^{(0,+)}$, indeed explains solar neutrino oscillations based on the LMA solution. The LMA solution is explicitly obtained by the use of the plausible parameter values for masses and couplings. The model has the following properties, which do not depend on the parameterization, that
\begin{enumerate}
   \item the global $L_e-L_\mu-L_\tau$ symmetry is explicitly broken by $\eta\eta S^{(+)c}S^{(+)c}$ down to a discrete $Z_{L^\prime}$ symmetry defined by $\exp(i\pi Q_{L^\prime}/2 )$ that forbids the $L$-breaking interactions for $S^{(+)}$, the absence of which is the essence of our discussions to be consistent with the bimaximal neutrino mixing scheme based on $L_e-L_\mu-L_\tau$;
   \item the $L$ conservation is explicitly broken by $\lambda_2\eta S^{(0)}\rho^c\chi^c$, which generates the VEV of $v^{(0)}_\nu$ for neutrinos calculated to be $\sim-\lambda_2v_\eta v_\rho/v_\chi$, where $\lambda_2 < 0$ is the trigger for $v^{(0)}_\nu \neq 0$;
   \item the mechanism generating the diagonal electron mass via $\delta m^\ell_{1i}$ ($i$=2,3) is linked to the requirement realizing the bimaximal neutrino mixing via $\delta m^\nu_{1i}$ owing to the proportionality of $\delta m^\nu_{1i}\propto\delta m^\ell_{1i}$; 
\end{enumerate}
and yields numerical results, which of course depend on the parameterization, that 
\begin{enumerate}
   \item the smallness of $m_{\nu 1}\approx m_{\nu 2}\sim 5.5 \times 10^{-2}$ eV for $\Delta m^2_{atm}$ is realized by $\vert\lambda_2\vert$ estimated to be ${\mathcal{O}}(10^{-7})$;
   \item solar neutrino oscillations are induced by 
   \begin{itemize}
   \item the one-loop radiative effects, which yield
        \begin{equation}
        \Delta m^2_{\odot}/\Delta m^2_{atm}\sim 10^{-2},
        \end{equation}
        corresponding to the LMA solution provided $SU(3)_L\times U(1)_N$ $\rightarrow$ $SU(2)_L\times U(1)_Y$ occurs at $3-6$ TeV,
   \item the effects of the simultaneous diagonalization of the neutrino and charged-lepton mass matrices, which are expected to yield
        \begin{equation}
        \Delta m^2_{\odot}/\Delta m^2_{atm} \mapleq (m_e/m_\mu)^{3/2} \sim 10^{-4},
        \end{equation}
        and which certainly yield the $U_{e3}$ matrix element estimated to be $\vert U_{e3}\vert \sim \sqrt{m_e/2 m_\mu} =0.05$,
   \end{itemize}
    where, in the present model, the dominance of one-loop radiative effects yields the LMA solution.  It should be stressed that the choice of $v_\chi$=${\mathcal{O}}$(1 TeV) automatically leads to the LMA solution.
\end{enumerate}

\begin{figure}
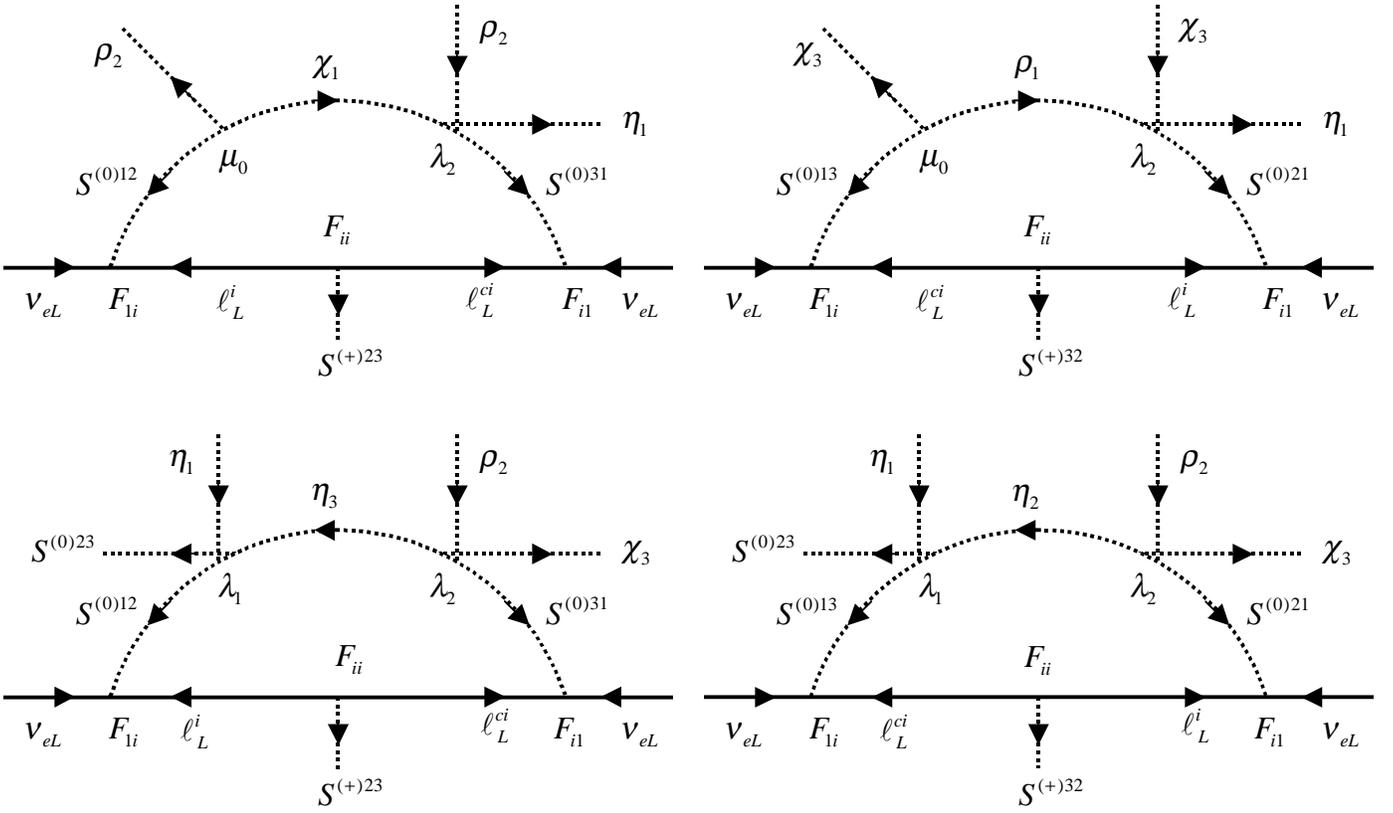

\caption{One-loop diagrams for  $\nu_e\nu_e$ with $\ell^i=\mu$ ($i$=2) and $\tau$ ($i$=3), where $m_{\ell^i}=F_{ii}v^{(+)}_\ell$ and $F_{1i}$ is the rotated $f_{1i}$-coupling defined by $F_{12} =\cos \alpha f_{12}  - \sin \alpha f_{13} $  and $F_{13} =\sin \alpha f_{12}  + \cos \alpha f_{13} $.}
\label{Fig:loopDiagrams_ee}
\end{figure}

\begin{figure}
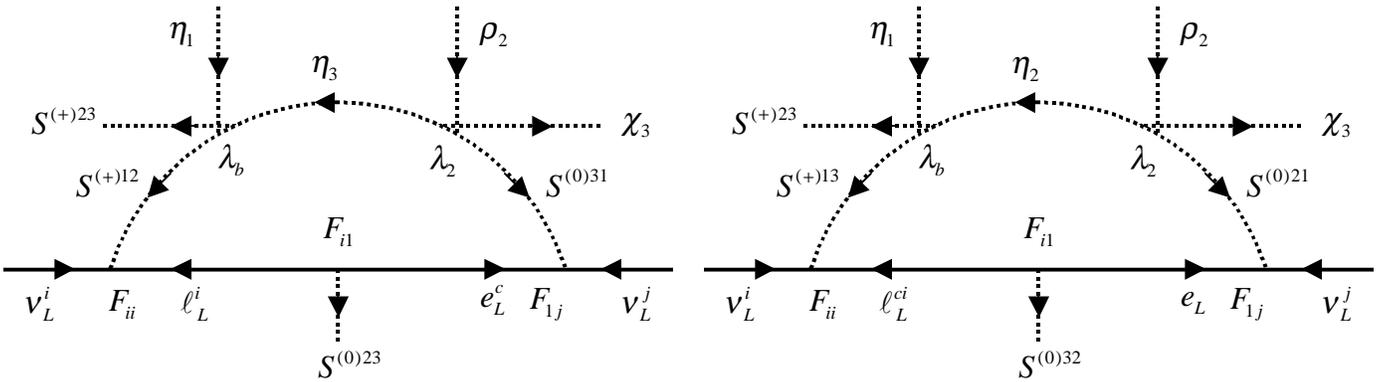

\caption{The same as FIG.\ref{Fig:loopDiagrams_ee} but for $\nu_{\mu,\tau}\nu_{\mu,\tau}$ with $\nu^{i,j}_L=\nu_{\mu L}$ ($i,j$=2) and $\nu_{\tau L}$ ($i,j$=3).}
\label{Fig:loopDiagrams_mutau}
\end{figure}

The present radiative mechanism generates the Majorana mass of $\nu_e$, $\delta m^{rad}_{11}$, given by the effective coupling of $(\eta^\dagger\psi^1)\psi^1\chi (\chi S^{(+)})^c$.  This coupling is induced by $\eta S^{(0)}\rho^c\chi^c$ and $\rho S^{(0)}\chi$ as in FIG.\ref{Fig:loopDiagrams_ee} and is of the $L^\prime$-conserving type.  The less-dominant radiative masses are induced by $\psi^i\psi^j(S^{(+)})^c(S^{(0)})^c\eta\rho\chi$ for $\nu_e\nu_e$ with $i,j$=1 contributing to $\delta m^{rad}_{11}$ and for $\nu_{\mu,\tau}\nu_{\mu,\tau}$ with $i,j$=2,3 contributing to $\delta m^{rad}_{ij}$.  The $\nu_{\mu,\tau}\nu_{\mu,\tau}$-terms involve the $L^\prime$-breaking coupling of $\eta\eta S^{(+)c}S^{(+)c}$ as in FIG.\ref{Fig:loopDiagrams_mutau}.  The splitting of $\nu_1$ and $\nu_2$ is provided by $\delta m^{rad}_{11}$, which is about $10^{-4}-10^{-3}$ eV, leading to the LMA solution.  The mass of $\nu_3$ is provided by $\delta m^{rad}_{33}$ at most of order $10^{-7}$ eV but the diagonalization due to $M_{\nu,\ell}$ may give $m_{\nu 3}$ $\mapleq$ $10^{-5}$ eV.

We have also presented plausible order-of-magnitude estimation of neutrino oscillations based on $M_\nu$, which is correlated to $M_\ell$:
\begin{enumerate} 
\item for the hierarchical texture with $\delta m^\nu_{12}\sim \delta m^\nu_{13}$,
\begin{eqnarray}\label{Eq:LeptonTextureH}
M_\nu=\epsilon_\nu
\left( 
\begin{array}{*{20}c}
   * & \sqrt{m_em_\mu} & \sqrt{m_em_\mu} \\
   \sqrt{m_em_\mu} & * & *\\
   \sqrt{m_em_\mu} & * & * \\
\end{array} 
\right),
M_\ell &=& 
\left( 
\begin{array}{*{20}c}
   0 & \sqrt{m_em_\mu} & \sqrt{m_em_\mu} \\
   \sqrt{m_em_\mu} & \delta M & \Delta M\\
   \sqrt{m_em_\mu} & \Delta M & M \\
\end{array}
\right)
,
\end{eqnarray}
\item for the democratic texture with $\delta m^\nu_{12}\gg \delta m^\nu_{13}$,
\begin{eqnarray}\label{Eq:LeptonTextureB1}
M_\nu=\epsilon_\nu
\left( 
\begin{array}{*{20}c}
   * & c_\phi\sqrt{m_em_\tau} & s_\phi\sqrt{m_em_\mu} \\
   c_\phi\sqrt{m_em_\tau} & * & *\\
   s_\phi\sqrt{m_em_\mu} & * & *\\
\end{array} 
\right),
M_\ell &=& 
\left( 
\begin{array}{*{20}c}
   0 & c_\phi\sqrt{m_em_\tau} & s_\phi\sqrt{m_em_\mu} \\
   c_\phi\sqrt{m_em_\tau} & M & M-\Delta M\\
   s_\phi\sqrt{m_em_\mu} & M-\Delta M & M \\
\end{array}
\right)
\end{eqnarray}
with $c_\phi=\cos\phi$ and $s_\phi=\sin\phi$ constrained by $\tan^2\phi \ll m_\tau/m_\mu$,
\item for the democratic texture with $\delta m^\nu_{12}\ll \delta m^\nu_{13}$,
\begin{eqnarray}\label{Eq:LeptonTextureB2}
M_\nu=\epsilon_\nu
\left( 
\begin{array}{*{20}c}
   * & \varepsilon_\ell\sqrt{m_em_\mu} & \sqrt{m_em_\mu} \\
   \varepsilon_\ell\sqrt{m_em_\mu} & * & *\\
   \sqrt{m_em_\mu} & * & * \\
\end{array} 
\right),
M_\ell &=& 
\left( 
\begin{array}{*{20}c}
   0 & \varepsilon_\ell\sqrt{m_em_\mu} & \sqrt{m_em_\mu} \\
   \varepsilon_\ell\sqrt{m_em_\mu} & M & M-\Delta M\\
   \sqrt{m_em_\mu} & M-\Delta M & M \\
\end{array}
\right)
\end{eqnarray}
with $\varepsilon_\ell \ll 1$,
\end{enumerate}
where $\epsilon_\nu$=$v^{(0)}_\nu/v^{(0)}_\ell$, $M\gg\Delta M,\delta M$ and the asterisks stand for the suppressed entries to generate $\Delta m^2_\odot$.  The partial $\ddash$degeneracy" between $M_\nu$ to $M_\ell$ suppresses the rotation effects on $\Delta m^2_{\odot}$ due to the diagonalization of $M_\nu$ and $M_\ell$.  As a result, the solar neutrino oscillations are, respectively, characterized by $\Delta m^2_{\odot}\sim (2.4, 6.9\cos\phi, 1.8) \times 10^{-5}$ eV$^2$ for $v_\chi\sim 4$ TeV.  Of course, if our mechanism explaining neutrino oscillations works further, one should next consider some reasons for the emergence of the approximate equality of $\delta m^\nu_{12} \sim \delta m^\nu_{13}$ in the hierarchical case and of the hierarchy of either $\delta m^\nu_{12} \gg \delta m^\nu_{13}$ or $\delta m^\nu_{12} \ll \delta m^\nu_{13}$ in the democratic case. 

The masses of Higgs scalars are simply assumed to be about $v_{weak}$ for those related to the weak boson masses for the order-of-magnitude estimate for radiative neutrino masses, namely, $m_{\eta^\pm}$ $\sim$ $m_{\rho^+}$ $\sim$ $m_{s^{(0)\pm}}$ $\sim$ $m_{s^{(+)\pm}}$ $\sim$ $v_{weak}$.  On the other hand, the $\chi$ scalar that generates masses for the exotic quarks and gauge bosons is assumed to have $m_{\chi^+}$ $\sim$ $v_\chi$($\gg$ $v_{weak}$) because $\langle 0 \vert \chi \vert 0 \rangle$ is related to the spontaneous breakdown of $SU(3)_L\times U(1)_N$ to $SU(2)_L\times U(1)_Y$.  We have estimated $v_\chi$ to lie in $3-6$ TeV in order to reproduce the observed neutrino mass pattern.  It is further known that the 331 model contains peculiar particles, {\it i.e.} doubly-charged bilepton gauge bosons \cite{Frampton} with masses of the order $ev_\chi/\sqrt{2}$, that couple to ${\overline {\ell^c_L}}\gamma_\mu\ell_L$. Since $v_\chi$ can be $3-6$ TeV, the bilepton gauge bosons can be as light as and even lighter than 1 TeV.  It is, therefore, quite conceivable that we will observe the bilepton gauge bosons with masses around 1 TeV, which provide striking phenomena with clean $\ell^-$-$\ell^-$ jets \cite{BiLepton}.

Now equipped with the plausible mechanism that generates phenomenologically consistent masses for neutrinos and charged leptons, the 331 model may be a physically interested model of leptons.  It not only explains three families of quarks and leptons because of the specific anomaly-cancellation mechanism used together with the asymptotic free condition on $SU(3)_c$ but also involves Majorana mass terms demanded by charged-lepton mass terms.  There remain dynamical questions about the smallness of the $L$-violating coupling of $\lambda_2$ and the fine tuning of the lepton mass parameters to reproduce $m_e\ll m_\mu \ll m_t$, which require some new physics beyond $SU(3)_L \times U(1)_L$. It is also required to have extensive analyses on physics of the quark and gauge boson sectors as well to fully clarify potential power of the 331 model.

\begin{center}
{\bf ACKNOWLEDGMENTS}
\end{center}

The work of M.Y. is supported by the Grant-in-Aid for Scientific Research No 12047223 from the Ministry of Education, Science, Sports and Culture, Japan.

\newpage
\vspace{10mm}
\noindent
\centerline{\bf Table Captions}
\\
\\
\centerline{TABLE \ref{Tab:Lnumber0}. $L^\prime$, $Z_4$ and $Z_{L^\prime}$ for leptons and Higgs scalars.}
\\
\\
\centerline{TABLE \ref{Tab:Lnumber}. $Z_4$ and $Z_{L^\prime}$ for non-self-Hermitian Higgs interactions.}
\\
\\
\noindent
\centerline{\bf Figure Captions}
\vspace{-6mm}
\vspace{5mm}
\begin{table}[ht]
    \caption{\label{Tab:Lnumber0} $L^\prime$, ${\mathcal{L}}$, $Z_4$ and $Z_{L^\prime}$ for leptons and Higgs scalars, where the lepton number $L$ is related to ${\mathcal{L}}$ as $L$ = $4\lambda^8/2\sqrt{3}$ + ${\mathcal{L}}$.}
    \begin{center}
    \begin{tabular}{lccccccc}\hline
           &$\psi^1$ & $\psi^{2,3}$ & $\eta$ & $\rho$ & $\chi$ & $S^{(0)}$ & $S^{(+)}$ 
\\ \hline
$L^\prime$    & 1    & $-1$         &  0     &   0    &   0    & 0         & 2
\\ \hline
${\mathcal{L}}$    &1/3   & 1/3          & $-$2/3 & $-$2/3 & 4/3    & $-$2/3    & $-$2/3
\\ \hline
$Z_4$         &+     & $-$          & +      & $i$    & $i$    & $-$       & $+$
\\ \hline
$Z_{L^\prime}$ & $i$ & $-i$        &  +      &  +     &  +     & +         & $-$
\\ \hline
    \end{tabular}
    \end{center}
\end{table}
\vspace{-10mm}
\begin{table}[ht]
    \caption{\label{Tab:Lnumber}$Z_4$ and $Z_{L^\prime}$ for non-self-Hermitian Higgs interactions.}
    \begin{center}
    \begin{tabular}{lcc|lcc|lcc|lcc}\hline
interactions&$Z_4$&$Z_{L^\prime}$&interactions&$Z_4$&$Z_{L^\prime}$&
interactions&$Z_4$&$Z_{L^\prime}$&interactions&$Z_4$&$Z_{L^\prime}$\\ \hline
$S^{(0)}S^{(0)}S^{(0)}$     &  $-$  &    +    &
$S^{(0)}S^{(0)}S^{(+)}$     &   +   &   $-$   &
$S^{(0)}S^{(+)}S^{(+)}$     &  $-$  &    +    &
$S^{(+)}S^{(+)}S^{(+)}$     &    +  &   $-$   \\

$\eta\eta S^{(0)c}S^{(0)c}$ &   +   &    +    &
$\eta\eta S^{(0)c}S^{(+)c}$ &  $-$  &   $-$   &
$\eta\eta S^{(+)c}S^{(+)c}$ &   +   &    +    &
$\rho\chi S^{(0)c}S^{(0)c}$ &  $-$  &    +    \\ 

$\rho\chi S^{(0)c}S^{(+)c}$ &   +   &   $-$   & 
$\rho\chi S^{(+)c}S^{(+)c}$ &  $-$  &    +    &
$\eta S^{(0)}\rho^c\chi^c$  &   +   &    +    & 
$\eta S^{(+)}\rho^c\chi^c$  &  $-$  &   $-$   \\ 

$\rho S^{(0)}\rho^c\eta^c$  &  $-$  &    +    &
$\rho S^{(+)}\rho^c\eta^c$  &   +   &   $-$   &
$\chi S^{(0)}\chi^c\eta^c$  &  $-$  &    +    &
$\chi S^{(+)}\chi^c\eta^c$  &   +   &   $-$   \\

$\eta S^{(0)}\eta$          &  $-$  &    +    &
$\eta S^{(+)}\eta$          &   +   &   $-$   &
$\rho S^{(0)}\chi$          &   +   &    +    &
$\rho S^{(+)}\chi$          &  $-$  &   $-$   \\

$\eta \rho \chi$            &  $-$  &    +    &
$(\rho^\dagger \eta)(\chi^\dagger \eta)$  &  $-$  &    +    &
Tr($S^{(0)\dagger}S^{(+)}$)& $-$ & $-$ & 
&& \\ \hline
    \end{tabular}
    \end{center}
\end{table}

\newpage
\centerline{\epsfbox{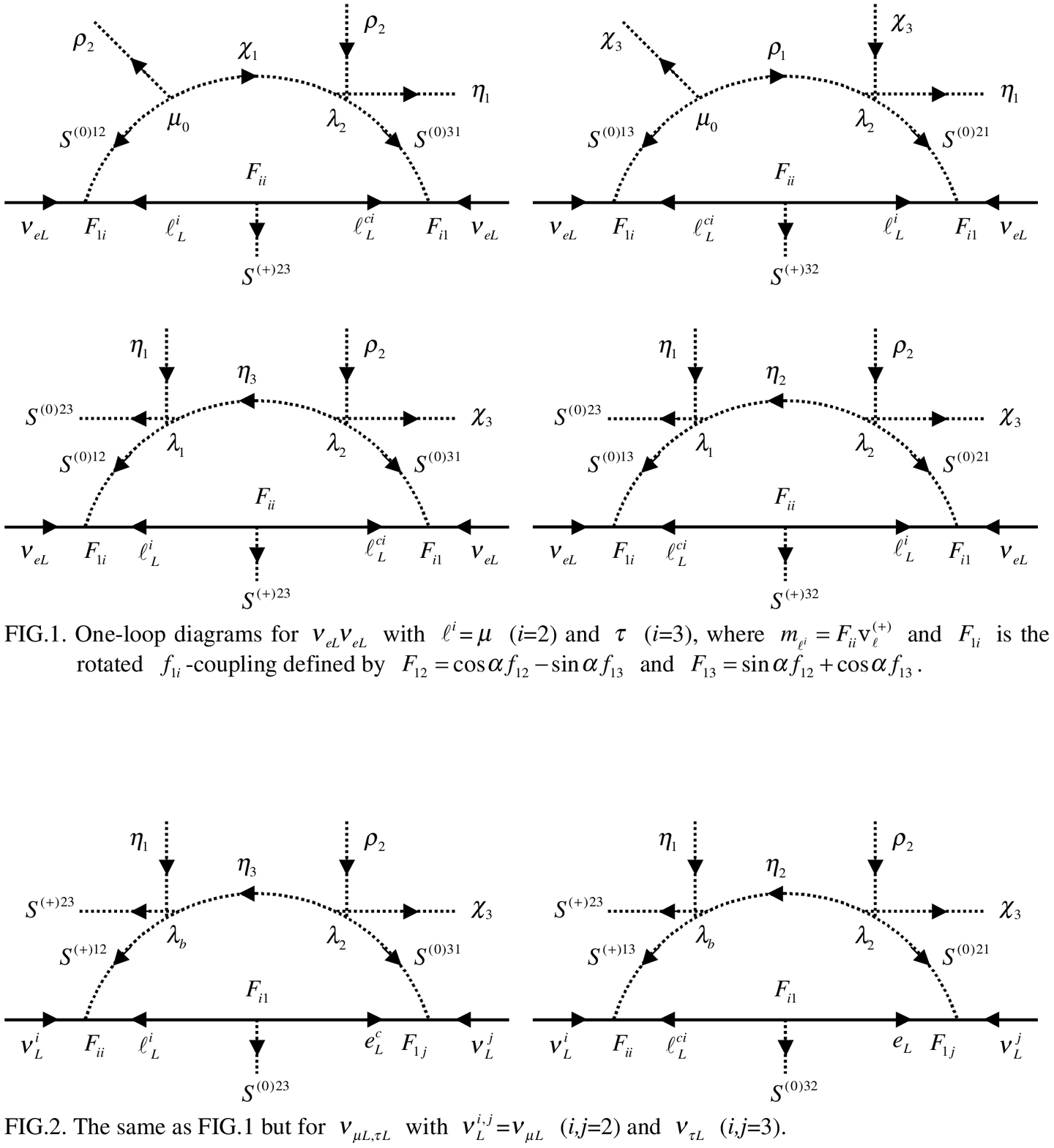}}

\end{document}